\documentclass[pre, twocolumn, showpacs,superscriptaddress]{revtex4}
\usepackage{amsmath}
\usepackage{graphicx}
\usepackage{bm}
\usepackage{ulem}
\usepackage{bm}

\begin{document}

\title{Supersymmetric Quantum Mechanics and Solitons of the sine-Gordon and Nonlinear Schr\"{o}dinger Equations}

\author{Andrew Koller}
\affiliation{Department of Physics, University of Massachusetts Boston, Boston MA 02125, USA}
\author{Maxim Olshanii}
\affiliation{Department of Physics, University of Massachusetts Boston, Boston MA 02125, USA}

\date{\today}

\begin{abstract}
We present a case demonstrating the connection between supersymmetric quantum mechanics (SUSY--QM), reflectionless scattering, and soliton solutions of integrable partial differential equations. We show that the members of a class of reflectionless Hamiltonians, namely, Akulin's
Hamiltonians, are connected via supersymmetric chains to a potential-free Hamiltonian, explaining their reflectionless nature. While the reflectionless 
property in question has been mentioned in the literature for over two decades, the enabling algebraic mechanism was previously 
unknown. Our results indicate that the multi-solition solutions of the sine-Gordon and nonlinear Schr\"{o}dinger equations can be systematically generated via the supersymmetric chains 
connecting Akulin's Hamiltonians. Our findings also explain a well-known but little-understood effect in laser physics: when
a two-level atom, initially in the ground state, is subjected to a laser pulse of the form
$V(t) = (n\hbar/\tau)/\cosh(t/\tau)$, with $n$ being an integer and $\tau$ being the pulse duration, 
it remains in the ground state after the pulse has been
applied, for {\it any} choice of the laser detuning.
\end{abstract}

\maketitle

\section{Introduction}
Reflectionless scattering, that is, scattering without reflection for {\it all} values of the spectral parameter, was first studied in the context of light propagation in a spatially inhomogeneous 
dielectric media \cite{kay1956_1503}. Mathematically, 
this is equivalent to finding
reflectionless potentials for 
the one-dimensional 
non-relativistic Schr\"{o}dinger equation \cite{shabat1992_303,spiridonov1992_398}. The underlying 
mechanism responsible for the reflectionless property of these potentials 
comes from the algebra of supersymmetric quantum mechanics (SUSY--QM) \cite{witten1981_513}, 
which links---via a finite number of intermediate steps---the reflectionless Hamiltonians to 
a potential-free Hamiltonian \cite{sukumar1985_2917,sukumar1986_2297,barclay1993_2786,cooper1988_1}.
Potential-free Hamiltonians are, in turn, inherently reflectionless.

It was later discovered that these same reflectionless quantum-mechanical potentials, when used as initial conditions for the Korteweg-deVries (KdV) equation, lead to multi-soliton solutions \cite{gardner1967_1095, scott1973_1443}. In particular, the reflectionless potential parameterized by an integer $n$ leads to an $n$-soliton solution of KdV. 

In this paper we present a SUSY--QM interpretation for a known class of reflectionless Hamiltonians
\cite{akulin2006_book, delone_1985} that lead to solitons of the sine-Gordon (sG) and attractive nonlinear Schr\"{o}dinger (NLS) equations. We also conjecture 
a connection between our SUSY chains and a class of Darboux transformations \cite{sall1982_1092,matveev1991_book}, thus 
allowing for a SUSY--QM explanation for the existence of the latter.

\section{Supersymmetric Quantum Mechanics}
Imagine we have two Hamiltonians $\hat{H}_0$ and $\hat{H}_1$, which are differential operators of finite order and dimension, whose coefficients may be functions of space 
but are asymptotically constant
as $x \to \pm \infty$, so that a scattering problem may be defined for them. 
$\hat{H}_0$ and $\hat{H}_1$ are known as supersymmetric partners if they can be factored as
\begin{eqnarray}
\label{susypartners}
\nonumber
\hat{H}_0 = \hat{\cal {A}}\hat{\cal {B}} + \epsilon; \quad
\hat{H}_1 = \hat{\cal {B}}\hat{\cal {A}} + \epsilon 
\quad,
\end{eqnarray}
where $\hat{\cal {A}}$ and $\hat{\cal {B}}$ are also differential operators, and $\epsilon$ is a constant, multiplied by the appropriate identity.
The operators $\hat{\cal {A}}$ and $\hat{\cal {B}}$ also 
intertwine the Hamiltonians:
\begin{eqnarray}
\nonumber
\hat{H}_1\hat{\cal {B}} = \hat{\cal {B}}\hat{H}_0; \qquad
\hat{H}_0\hat{\cal {A}} = \hat{\cal {A}}\hat{H}_1
\quad.
\end{eqnarray}
Eigenstates of  $\hat{H}_0$ and $\hat{H}_1$ are related by 
\begin{eqnarray}
\nonumber
\hat{\cal {B}}|\psi_0\rangle \propto |\psi_1\rangle; \quad
\hat{\cal {A}}|\psi_1\rangle \propto |\psi_0\rangle 
\quad,
\end{eqnarray}
where $|\psi_0\rangle$ is an eigenstate of $\hat{H}_0$ and $|\psi_1\rangle$ is an eigenstate of $\hat{H}_1$ with the same eigenvalue \cite{sukumar1985_2917}. 
It is
also
possible to have a hierarchy, or chain, of such supersymmetric partners:
\begin{eqnarray}
\label{susychain}
\hat{H}_0 &=& \hat{\cal {A}}_1\hat{\cal {B}}_1 + \epsilon_1\\
\nonumber
\hat{H}_1 &=& \hat{\cal {B}}_1\hat{\cal {A}}_1 + \epsilon_1 = \hat{\cal {A}}_2\hat{\cal {B}}_2 + \epsilon_2\\
\nonumber
\vdots\\
\nonumber
\hat{H}_m &=& \hat{\cal {B}}_m\hat{\cal {A}}_m+\epsilon_m = \hat{\cal {A}}_{m+1}\hat{\cal {B}}_{m+1} + \epsilon_{m+1}\\
\nonumber
\vdots\\
\nonumber
\hat{H}_n &=& \hat{\cal {B}}_{n}\hat{\cal {A}}_{n}+\epsilon_{n} = \dots
\quad.
\end{eqnarray}
In the case of a SUSY--QM chain, eigenstates of $\hat{H}_n$ will be linked to eigenstates of $\hat{H}_m$ as
\begin{eqnarray}
\label{mn}
|\psi_n\rangle \propto  \hat{\Upsilon}_{n\leftarrow m} |\psi_m\rangle
\quad,
\end{eqnarray}
where $\hat{\Upsilon}_{n\leftarrow m} = \hat{\cal {B}}_{n}\hat{\cal {B}}_{n-1}\ldots\hat{\cal {B}}_{m}$ in this particular case. More generally,
the operator $\hat{\Upsilon}_{n_{2}\leftarrow n_{1}}$ constitutes a so-called {\it intertwiner} that relates Hamiltonian a $\hat{H}_{n_{1}}$ to a Hamiltonian $\hat{H}_{n_{2}}$. We define an intertwiner 
as an operator that 
obeys the following property: 
\begin{eqnarray}
\label{defineintertwiner}
&&
\hat{H}_{n_{2}} \hat{\Upsilon}_{n_{2}\, \leftarrow\, n_{1}} = \hat{\Upsilon}_{n_{2}\, \leftarrow\, n_{1}} \hat{H}_{n_{1}}
\quad,
\end{eqnarray}
which guarantees that $\hat{\Upsilon}_{n_{2}\, \leftarrow\, n_{1}}$ acting on an eigenstate of
$\hat{H}_{n_{1}}$ produces a linear combination of the eigenstates of $\hat{H}_{n_{2}}$ of the same energy.
The intertwiners can be nested  as
$
\hat{\Upsilon}_{n_{2}\, \leftarrow\, n_{1}} =
\hat{\Upsilon}_{n_{2}\, \leftarrow\, n'}\hat{\Upsilon}_{n'\, \leftarrow\, n_{1}} =
\hat{\Upsilon}_{n_{2}\, \leftarrow\, n''}\hat{\Upsilon}_{n''\, \leftarrow\, n'}\hat{\Upsilon}_{n'\, \leftarrow\, n_{1}} =
\ldots
$.
In particular: $\hat{\Upsilon}_{n\, \leftarrow\, m}=\hat{\Upsilon}_{n\, \leftarrow\, n-1}\hat{\Upsilon}_{n-1\, \leftarrow\, n-2}\ldots\hat{\Upsilon}_{m+1\, \leftarrow\, m}$. Note that in our case
$\hat{\Upsilon}_{l\, \leftarrow\, l-1} = \hat{\cal {B}}_{l}$, hence the expression for $\hat{\Upsilon}_{n\, \leftarrow\, m}$ above.

Of particular interest is the case in which one of the members of the chain, say $\hat{H}_0$, is ``potential-free" in the sense that it is a differential operator
whose coefficients are constant everywhere.
$\hat{H}_0$ is inherently reflectionless since its eigenstates are just plane waves. If the coefficients in the
$\hat{\Upsilon}_{l\leftarrow l-1}$ are asymptotically constant,
relationship \eqref{mn}, considered at $m=0$, implies that every other member of the supersymmetric chain \eqref{susychain} is also reflectionless. In general, Hamiltonians are reflectionless if they are connected to a potential-free Hamiltonian via a supersymmetric chain.

\section{Akulin's Hamiltonians}
We studied a class of $2\times2$ matrix differential operators given by
\begin{eqnarray}
&&
\hat{H}_{n} =  \sigma_{z} \partial_{x}  -  \sigma_{x} n/\cosh(x) 
\quad,
\label{H_n}
\\
&&
n = \ldots,\,-3,\,-2,\,-1,\,0,\,+1,\,+2,\,+3,\,\ldots
\quad,
\nonumber
\end{eqnarray}
which we refer to as Akulin's Hamiltonians 
(see pp.\ 208-210 in \cite{akulin2006_book}). 
We found a (non-unique) supersymmetric chain linking all the Hamiltonians of form \eqref{H_n}. Since $\hat{H}_0$ is potential-free and inherently reflectionless, every other member of the chain is also reflectionless. These Hamiltonians are not directly linked, however; there is an intermediate Hamiltonian, which we refer to as $\hat{H}_{n+1/2}$, between $\hat{H}_n$ and $\hat{H}_{n+1}$. The Hamiltonians $\hat{H}_{n+1/2}$ do not appear to have any physical significance and furthermore, they contain spatial-dependencies that diverge as $x \to \pm \infty$, making the definition of a scattering problem impossible.

\begin{figure}
\centering
\includegraphics[scale=0.35]{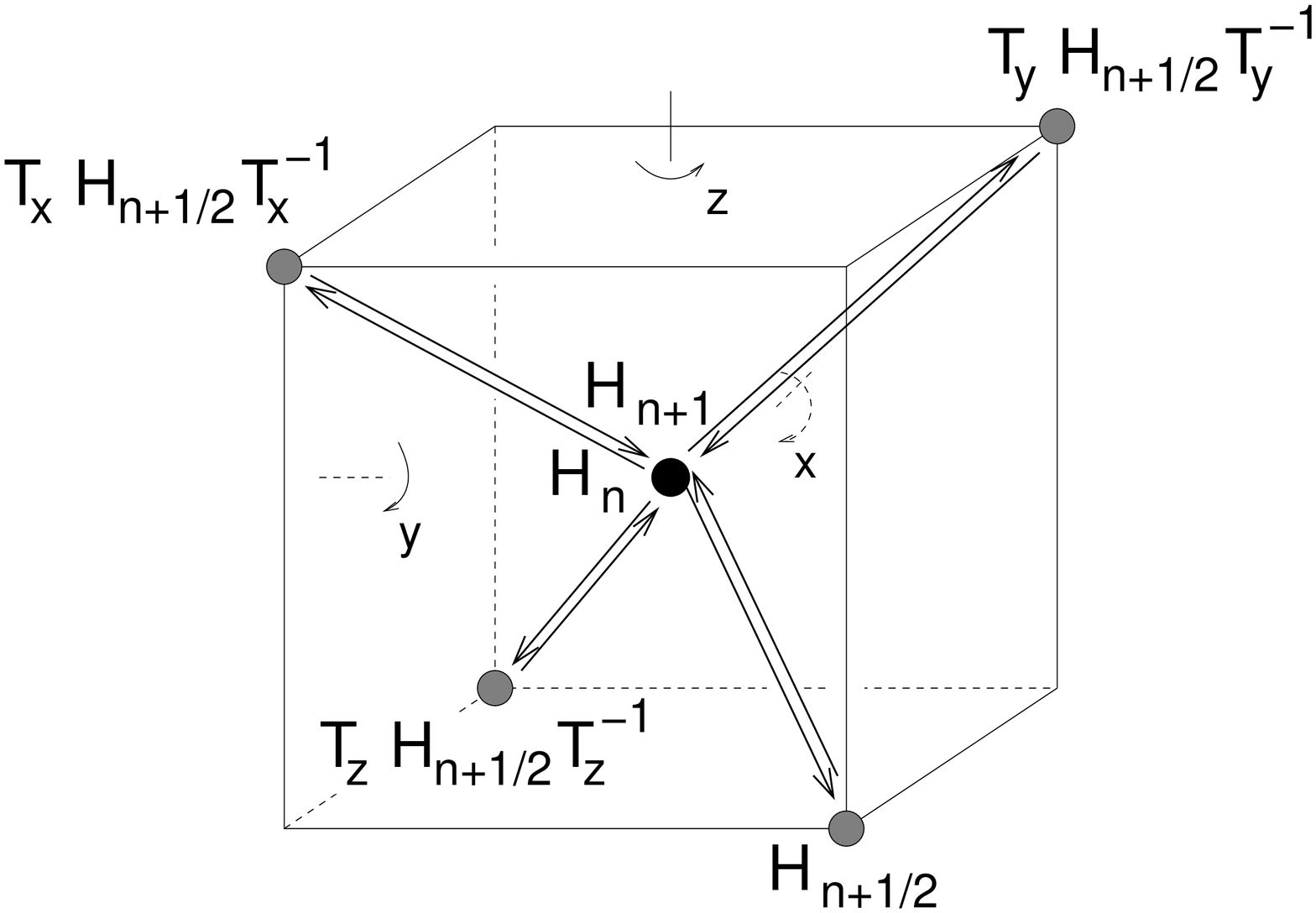}
\caption[Four SUSY--QM chains for Akulin's Hamiltonians]{Four SUSY--QM chains for Akulin's Hamiltonians: thick arrows correspond to SUSY--QM connections. The $180^{\circ}$ rotations about the axes $OX$,
$OY$, and $OZ$ correspond to the transformations ${\cal T}_{x} \equiv \hat{R} \sigma_{x} \cdot \bigcirc \cdot \sigma_{x}^{-1} \hat{R}^{-1}$, 
${\cal T}_{y} \equiv (i \sigma_{x}) \cdot \bigcirc \cdot (i \sigma_{y})^{-1}$, 
and ${\cal T}_{z} \equiv \hat{R} \sigma_{z} \cdot \bigcirc \cdot \sigma_{z}^{-1} \hat{R}^{-1}$ 
respectively. 
}
\label{fourchains}
\end{figure}

We used several symmetries of the Hamiltonians $\hat{H}_n$ to remove some ambiguities in defining such a chain. Consider the following transformations of {\it operators}:
$
{\cal T}_{I} \equiv \bigcirc; \quad
{\cal T}_{x} \equiv \hat{R} \sigma_{x} \cdot \bigcirc \cdot \sigma_{x}^{-1} \hat{R}^{-1}; \quad
{\cal T}_{y} \equiv (i \sigma_{y}) \cdot \bigcirc \cdot (i \sigma_{y})^{-1}; \quad
{\cal T}_{z} \equiv \hat{R} \sigma_{z} \cdot \bigcirc \cdot \sigma_{z}^{-1} \hat{R}^{-1},
$
where $\hat{R}$ refers to the operation $x \to -x$, and 
$\bigcirc$ denotes a slot for the operator on which the transformation is applied.
Under these actions, the Hamiltonians transform simply:
$
\hat{H}_{n} \stackrel{{\cal T}_{x}}{\leftrightarrow} +\hat{H}_{n}; \quad
\hat{H}_{n} \stackrel{{\cal T}_{y}}{\leftrightarrow} -\hat{H}_{n}; \quad
\hat{H}_{n} \stackrel{{\cal T}_{z}}{\leftrightarrow} -\hat{H}_{n},
$
forming a one-dimensional representation of the group $\mbox{Dih}_{2} = Z_{2} \times Z_{2}$, 
corresponding to $180^{\circ}$ rotations about 
coordinate axes, plus an identity. 
As a result, at every step of the SUSY--QM chain, the chain quadrifurcates as 
\begin{eqnarray}
\label{akulinchain}
\begin{array}{ccccccccc}
\hat{H}_{n} &=& \ldots = s_{{\cal T}} \left( {\cal T}[\hat{B}_{n}^{(+)}] {\cal T}[\hat{A}_{n}^{(+)}] +  \epsilon_{n}^{(+)}\right)\\
\hat{H}_{n+1/2}       
      &=& s_{{\cal T}} \left( {\cal T}[\hat{A}_{n}^{(+)}]   {\cal T}[\hat{B}_{n}^{(+)}]   +  \epsilon_{n}^{(+)}\right)\\
      &=& s_{{\cal T}} \left( {\cal T}[\hat{A}_{n+1}^{(-)}] {\cal T}[\hat{B}_{n+1}^{(-)}] +  \epsilon_{n+1}^{(-)}\right)
      & &
\\
\hat{H}_{n+1}       
      &=& s_{{\cal T}} \left({\cal T}[\hat{B}_{n+1}^{(-)}] {\cal T}[\hat{A}_{n+1}^{(-)}] +  \epsilon_{n+1}^{(-)}\right) 
      &=& \ldots
\end{array}
\end{eqnarray}
where $s_{{\cal T}} = {\cal T}[\hat{H}]/\hat{H}$, ${\cal T}$ can be chosen from $\{{\cal T}_{I},\,{\cal T}_{x},\,{\cal T}_{y},\,{\cal T}_{z} \}$, and the $s$ values are $s_I = +1$, $s_x = +1$, $s_y = -1$, and $s_z = -1$.
The base SUSY--QM factors are
\begin{widetext}
\begin{eqnarray}
\label{As_and_Bs}
\nonumber
\hat{B}_{n}^{(+)} &=& 
\left(
\begin{array}{cc}
1                                                                      &  -\frac{1}{2} n/\cosh(x)
\\
(-1)^{n}\cosh(x) - \sinh(x)                                            &  -\partial_{x}  -\frac{1}{2} \left( (-1)^{n} (n+1) + n\tanh(x) \right)
\end{array}
\right)
\\
\hat{A}_{n}^{(+)} &=& 
\left(
\begin{array}{cc}
+\partial_{x} -\frac{1}{2} \left( (-1)^{n} (n+1) + n\tanh(x) \right)    &  -\frac{1}{2} n/\cosh(x) 
\\
(-1)^{n}\cosh(x) - \sinh(x)                                            &  1
\end{array}
\right)
\nonumber
\\
\hat{A}_{n}^{(-)} &=&
(-1) \times
\left(
\begin{array}{cc}
-\partial_{x} +\frac{1}{2} \left( (-1)^{n} (n-1) - n\tanh(x) \right)    &  +\frac{1}{2} n/\cosh(x)
\\
(-1)^{n}\cosh(x) + \sinh(x)                                            &  1
\end{array}
\right)
\nonumber
\\
\hat{B}_{n}^{(-)} &=&
\left(
\begin{array}{cc}
1                                                                      &  +\frac{1}{2} n/\cosh(x)
\\
(-1)^{n}\cosh(x) + \sinh(x)                                            &  \partial_{x} + \frac{1}{2} \left( (-1)^{n} (n-1) - n\tanh(x) \right)
\end{array}
\right)
\quad,
\nonumber
\end{eqnarray}
\end{widetext}
with factorization constants
\begin{eqnarray}
\epsilon_{n}^{(+)} = (-1)^{n} (n+\frac{1}{2}); \quad
\epsilon_{n}^{(-)} = (-1)^{n} (n-\frac{1}{2})
\quad. \nonumber
\end{eqnarray}
Note that different ${\cal T}$ produce completely different intermediate Hamiltonians $\hat{H}_{n+1/2}$.
The four two-link 
SUSY--QM chains connecting $\hat{H}_n$ to $\hat{H}_{n+1}$ via different intermediate Hamiltonians $\hat{H}_{n+1/2}$ are illustrated in Fig. \ref{fourchains}. 

The intertwiners linking eigenstates of $\hat{H}_n$ to eigenstates of $\hat{H}_{n+1}$ $(\hat{\Upsilon}_{n+1\, \leftarrow\, n})$, and 
the intertwiners linking eigenstates of $\hat{H}_n$ to eigenstates of $\hat{H}_{n-1}$ $(\hat{\Upsilon}_{n-1\, \leftarrow\, n})$ 
can be easily extracted from the chain \eqref{akulinchain}:  
$
\hat{\Upsilon}_{n+1\, \leftarrow\, n}  = v_{{\cal T}} \left( {\cal T}[\hat{B}_{n+1}^{(-)}] {\cal T}[\hat{A}_{n}^{(+)}] \right)$, and
$\hat{\Upsilon}_{n-1\, \leftarrow\, n}  = v_{{\cal T}} \left( {\cal T}[\hat{B}_{n-1}^{(+)}] {\cal T}[\hat{A}_{n}^{(-)}] \right)$,
where we fix the arbitrary phase factor $v_{{\cal T}}$ to
$
v_{{\cal T}} = {\cal T}[\hat{\Upsilon}]/\hat{\Upsilon}
$, with values $v_I = +1$, $v_x = -1$, $v_y = +1$, and $v_z = -1$.
Substituting the expressions for $\hat{A}$'s given above, one obtains the following expressions 
for the intertwiners:
\begin{eqnarray}
&&
\left.\right.\hspace{-35pt}
\hat{\Upsilon}_{n+1\, \leftarrow\, n} =  \partial_{x}  -  (n+\frac{1}{2})\tanh(x)  + \frac{1}{2} (i\sigma_{y})/\cosh(x)                                
\nonumber
\\
&&
\left.\right.\hspace{-35pt}
\hat{\Upsilon}_{n-1\, \leftarrow\, n} =  \partial_{x}  +  (n-\frac{1}{2})\tanh(x)  - \frac{1}{2} (i\sigma_{y})/\cosh(x)                                
\label{intertwiners}
.
\end{eqnarray}
Interestingly, while in the intermediate expressions, the intertwiners depend on ${\cal T}$, this dependence disappears 
in the final expressions \eqref{intertwiners}.  
Thus, like Hamiltonians $\hat{H}_n$, the intertwiners form a one-dimensional representation of the group $\mathrm{Dih}_{2}$.

This (unexplained) property is nontrivial, because the eigenstates of $\hat{H}_n$ are doubly-degenerate, and thus, generally, different versions of the intertwiners can target different linear combinations 
of the eigenstates. 

\section{Some extensions of the class of reflectionless Hamiltonians}
So far, we have shown that Hamiltonians 
$
\hat{H}_{n} = \sigma_z\partial_x - \sigma_x w_{n}(x)
$
are reflectionless for $w_{n}(x) = n/\cosh(x)$, for all integer values of $n$. 
Here, $\sigma_{x,y,z}$ are Pauli
matrices. 
It can be easily shown that 
the class of reflectionless Hamiltonians can be extended to  
\begin{eqnarray}
&&
\hat{H}_{n}(\xi,\,\eta,\,x_{0},\,\phi) =
\nonumber
\\
&&\qquad\qquad
\sigma_z\partial_x - \frac{1}{2}(\sigma_{+} w_{n}(x;\,\xi,\,\eta,\,x_{0},\,\phi) + h.c.)
\nonumber
\\
&&
w_{n}(x;\,\xi,\,\eta,\,x_{0},\,\phi) = \frac{n \xi \exp[i (\eta (x-x_{0}) + \phi) ]}{\cosh[\xi (x-x_{0})]}
\,,
\label{H_n_BIS}
\end{eqnarray}
where $\sigma_{+} = \sigma_{x} + i \sigma_{y}$.

\section{Inverse Scattering Method}
The inverse scattering method was developed to solve the initial value problem for integrable nonlinear partial differential 
equations (NPDE's). Associated with each integrable NPDE are two linear differential operators, 
$\hat{L}=\hat{L}(u)$ and $\hat{M}=\hat{M}(u)$, known as the Lax pair. 
The solution $u(x,t)$ of the NPDE appears as a necessary condition for  
$\hat{L}$ and $\hat{M}$ to satisfy a Heisenberg-type equation $\dot{\hat{L}} = [\hat{M}, \hat{L}]$
\cite{Fordy_solitons, solitons_book}. Trivially related to $\hat{L}$ is a Hamiltonian $\hat{H}$ that 
defines a spectral problem $\hat{H}\psi = \lambda\psi$ \cite{ablowitz1973_125, solitons_book}.
In the case of KdV, $\hat{L} = \hat{H}$. For the sine-Gordon and attractive nonlinear Schr\"{o}dinger equations, 
$\hat{L} = \hat{\sigma}_z(\hat{H}-\lambda)$. The eigenvector $\psi$ evolves in time through $\hat{M}$.

The procedure for finding $u(x,t)$ is as follows. First, find the scattering data at $t = 0$, $S(0)$, for $\hat{H}(u(x,0))$.
Next, use the time evolution through $\hat{M}$ to find the scattering data at time $t$, $S(t)$.
Lastly, invert the scattering data at time $t$ to find $u(x,t)$. 
If the $\hat{H}$ operator for a NPDE is reflectionless, then the inversion formula, known as the 
Gel'fand-Levitan-Marchenko  
integral equation, greatly simplifies \cite{solitons_book, Fordy_solitons}, reducing to a system 
of {\it algebraic} equations, and leads to soliton solutions of the NPDE. 
Thus, finding SUSY--QM chains of reflectionless Hamiltonians becomes a method of generating large 
families of multi-soliton solutions of the corresponding NPDE.

\section{Solitons of the sine-Gordon equation}
The sine-Gordon equation is given by 
\begin{eqnarray}
\frac{\partial^2}{\partial x \partial t}u = \sin(u)
\quad,\nonumber
\end{eqnarray}
where $u = u(x,t)$, and $x$ and $t$ are light-cone coordinates, related to lab-frame coordinates $\bar{x}$ and $\bar{t}$ by $x = (\bar{x} + \bar{t})/{2}$, $t = (\bar{x} - \bar{t})/{2}$. Direct-scattering for sG is done via:
\begin{equation}
\hat{H}_{SG} = \sigma_z\partial_x - \sigma_x q(x,t=0)
\quad,
\label{H_SG}
\end{equation}
where $q(x,t) = -u_x(x,t)/2$. For $\eta = \phi_{0} = 0$,  
Akulin's Hamiltonians \eqref{H_n_BIS} have this same structure, and they give reflectionless direct-scattering initial conditions for the sine-Gordon equation. 
Several known solitonic solutions of sG can be identified. In particular, $\hat{H}_{\mp 1}(\xi \geq 0,\,0,\,x_{0},\,0)$ leads to a kink(anti-kink) soliton of initial position $x_{0}$ and
velocity $v=(1-\xi^2)/(1+\xi^2)$
in the lab frame:
\begin{equation}
u(x,t) = -4\tan^{-1}\bm{\left(}\exp\left\{\mp [\xi (x-x_{0}) + t/\xi]\right\}\bm{\right)}
\quad.\nonumber
\end{equation}

\begin{figure}
\centering
\includegraphics[scale=0.35]{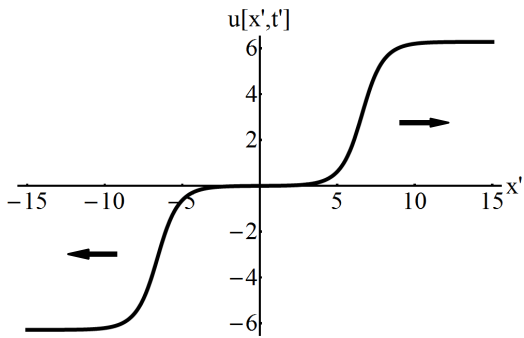}
\caption[2-soliton Solutions of sine-Gordon equation]{2-soliton solution of sine-Gordon equation: scattering with Hamiltonian $\hat{H}_{- 2}(\xi = 1/\sqrt{3},\,0,\,0,\,0)$ 
leads to a two-soliton solution where two kinks, initially located at the origin, move apart with speeds $v = \pm 1/2$ in the stationary lab frame, shown above. 
Other values of $\xi$ yield the same solutions but viewed from moving frames. 
}
\label{solitons}
\end{figure}

The Hamiltonian
$\hat{H}_{\mp 2}(\xi \geq 0,\,0,\,x_{0},\,0) $ leads to a two-soliton solution,
\begin{equation}
u(x,t) = -4\tan^{-1}\bm{\left(}\frac{\mp \sinh\left\{2[\xi (x-x_{0}) + t/3\xi]\right\}}{2\cosh\left[-\xi (x-x_{0}) + t/3\xi\right]}\bm{\right)}
\quad,\nonumber
\end{equation}
which consists of two kinks(anti-kinks) that, at $t=0$ 
collide at $x_{0}$ with
velocities $v_{1}=+1/2$ and $v_{2}=-1/2$ in the lab frame (Fig. \ref{solitons}), and that are 
observed from a reference frame of velocity
$V=(3\xi^2-1)/(3\xi^2+1)$ with respect to the lab frame.
%

\section{Connection between the SUSY intertwiners and Darboux transformations}
At this point, it is instructive to compare the intertwiners (\ref{intertwiners}) to the (well-known) Darboux transformations 
\cite{sall1982_1092,matveev1991_book},
that also allow the generation of new soliton solutions of 
nonlinear equations from known solutions. 
In the context of the sine-Gordon equation, the Darboux connection states, in particular, that 
for any eigenstate $|\psi(\lambda)\rangle$ of the Hamiltonian $\hat{H}_{SG}$ (\ref{H_SG}) (parametrized by some $q(x)$) corresponding 
to an eigenvalue $\lambda$, the state 
\begin{eqnarray}
|\tilde{\psi}(\lambda)\rangle = \hat{U} |\psi(\lambda)\rangle
\nonumber
\end{eqnarray}
is an eigenstate of a new Hamiltonian $\hat{\tilde{H}}_{SG}$ (parametrized by $\tilde{q}(x)$) 
with the same eigenvalue. The intertwining operator $\hat{U}$ has the form   
\begin{eqnarray}
&&
\hat{U} = 
- \frac{1}{2} \lambda_{0} (\zeta(x)+\frac{1}{\zeta(x)})
\nonumber
\\
&&
\quad
- \frac{1}{2} \lambda_{0} (\zeta(x)-\frac{1}{\zeta(x)}) \sigma_y
+
\lambda \sigma_z
\quad,
\label{Darboux_intertwiner}
\end{eqnarray}
and the new field $q$ is 
\begin{eqnarray}
\tilde{q}(x) = -q(x,t) + \lambda_{0} (\zeta(x)-1/\zeta(x)) )
\quad,
\nonumber
\end{eqnarray}
where $\zeta(x) = (\varphi_{1}(x)+i \varphi_{2}(x))/(\varphi_{1}(x)-i \varphi_{2}(x))$, and 
$(\varphi_{1}(x),\varphi_{2}(x))^{T} = |\phi(\lambda_{0})\rangle$ forms a ``fixed'' eigenstate of $\hat{H}_{SG}$ of an 
eigenvalue $\lambda_{0}$. We conjecture that for any intertwiner $\hat{\Upsilon}$ (\ref{intertwiners}) identified above,
there exists a Darboux operator $\hat{U}$ (\ref{Darboux_intertwiner}), such that $\hat{\Upsilon}$ and $\hat{U}$ 
coincide on the subspace of the eigenstates of $\hat{H}_{SG}$ of eigenvalue $\lambda$. For example, 
the choice of $\lambda_{0}=1/2$ and $|\phi(\lambda_{0})\rangle = (\exp[+x/2],\exp[-x/2])^{T}$ leads 
to the intertwiner $\hat{\Upsilon}_{+1\, \leftarrow\, 0}$ (\ref{intertwiners}). 

According to our conjecture, 
our intertwiners do {\it not} generate new, previously unknown solutions of the sine-Gordon equation. 
Nevertheless, the intertwiners (\ref{intertwiners}) allow one to interpret the Darboux links between various Hamiltonians $\hat{H}_{SG}$ \eqref{H_SG} 
as a consequence a hidden SUSY--QM structure. Recall that in the case of the KdV equation (whose analog of $\hat{H}_{SG}$ 
is the usual stationary linear Schr\"{o}dinger 
operator), a similar interpretation is well known    
\cite{sukumar1985_2917,sukumar1986_2297,barclay1993_2786,cooper1988_1}. At the same time, we know of no published work which finds a
similarly simple explanation for the existence of Darboux intertwiners \eqref{Darboux_intertwiner}.

\section{Solitons of the Nonlinear Schr\"{o}dinger equation}
The nonlinear Schr\"{o}dinger equation with attractive interactions is given by
\begin{eqnarray}
i\frac{\partial}{\partial t}u = -\frac{\partial^2}{\partial x^2}u - 2|u|^2 u
\quad.\nonumber
\end{eqnarray}
Direct-scattering is done with
\begin{equation}
\hat{H}_{NLS} = 
\sigma_z\partial_x - \frac{1}{2}(\sigma_{+} q(x,t=0) + h.c.)
\quad,\nonumber
\label{Hnls}
\end{equation}
with $q(x,t) = u(x,t)$. Akulin's Hamiltonians \eqref{H_n_BIS}, $\hat{H}_{\pm 1}(\xi>0,\,\eta,\,x_{0},\,\phi)$ produce 
all one-soliton solutions:
\begin{eqnarray}
u(x,t) = \pm \xi \exp[i (v x'/2+\phi)] \, \frac{\exp[i t']}{\cosh[x']}
\quad,\nonumber
\end{eqnarray}
where $v = 2\eta$ is the soliton velocity, $x_{0}$ is its initial position, $\phi$ is its phase \cite{solitons_book},
$x'=\xi (x-vt -x_{0})$, and $t = \xi^2 t$.
In turn, $\hat{H}_{\pm 2}(\xi>0,\,\eta,\,x_{0},\,\phi)$ generates the following breather-like two-soliton solutions:
\begin{eqnarray}
&&
u(x,t) = \pm \xi \exp[i (v x'/2+\phi)]
\nonumber
\\
&&
\qquad
\times
\frac{4 e^{i t'} \left(\cosh (3 x')+3 e^{8 i t'} \cosh (x')\right)}{3 \cos (8 t')+4 \cosh (2 x')+\cosh (4 x')}
\quad,\nonumber
\end{eqnarray}
where $v$, $x_{0}$, and $\phi$ now correspond to the velocity, initial position, and the overall phase of the breather, respectively. 
All $n$-soliton solutions of this type were first obtained, using the inverse scattering method,
in Ref. \cite{schrader1995_2221}. Note that in the NLS case, as well as in the sG case, 
multisoliton solutions can be generated via Darboux transformations \cite{sall1982_1092,matveev1991_book}.

\section{The {\rm sech}-shaped Laser Pulses}
\begin{figure}
\centering
\includegraphics[scale=0.35]{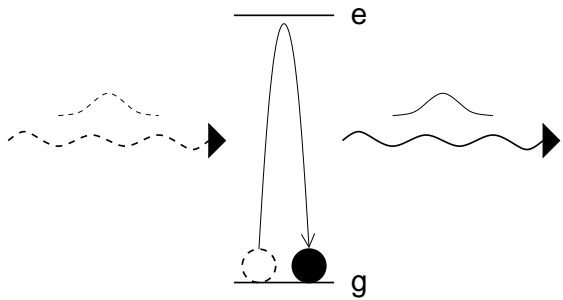}
\caption[Inversionless Laser Pulse]{{\rm Sech}-shaped laser pulse: our SUSY--QM connection of Akulin's Hamiltonians to the potential-free $\hat{H}_0$ also explains 
a series of specific laser pulse shapes that exhibit 
a zero ground-excited population 
transfer for {\it any} value of the laser detuning.
}
\label{pulse}
\end{figure}

Akulin's Hamiltonians were first studied by Akulin in the context of a two-level, time-dependent system \cite{akulin2006_book} that maps to the spatial scattering problem we have considered in this paper: here we present the original problem. Consider a two-level atom subjected to a time-dependent pulse of the form $V_{eg}(t) = V/\cosh\left(t/\tau\right)$
and detuning $\Delta$.
Here $V$ is the amplitude of the pulse, $\tau$ is its duration, and $|e\rangle$ and $|g\rangle$ are the excited and ground states, respectively. 
If we represent the probability amplitudes of the
ground and excited states by $\psi_{g}$ and $\psi_{e}$, respectively, the dynamics of the system will obey
\begin{eqnarray}
\nonumber
i\frac{d}{d t}\psi_{g} &=& +\frac{\Delta}{2} \psi_{g} + \frac{n}{\tau\cosh\left(t/\tau\right)}\psi_{e} 
\\
i\frac{d}{d t}\psi_{e} &=& +\frac{n}{\tau\cosh\left(t/\tau\right)} \psi_{g} - \frac{\Delta}{2} \psi_{e}
\quad. \nonumber
\label{H_lp}
\end{eqnarray}
It is known that 
for specific values of the pulse amplitude, given by $V = n\hbar/\tau$, where $n$ is an integer,
the transition probability is zero {\it regardless} of the detuning choice $\Delta$ \cite{akulin2006_book} (Fig. \ref{pulse}). 
Mathematically, this property was not well understood.
However, a re-interpretation of this problem as finding eigenstates of a reflectionless Akulin's Hamiltonian 
$\hat{H}_{n}(\xi=1/\tau,\,0,\,0,\,\pi/2)$ \eqref{H_n_BIS} with an eigenvalue $\lambda= -i \Delta/2$ explains 
the absence of the population transfer in the pulse propagation. 
%

\section{Summary and Outlook}
We have shown that Akulin's Hamiltonians \eqref{H_n} are connected to a potential-free Hamiltonian via supersymmetric chains \eqref{akulinchain}, 
explaining their reflectionless nature. Akulin's Hamiltonians lead to multi-soliton solutions when used as direct-scattering initial conditions for the
sine-Gordon and attractive nonlinear Schr\"{o}dinger equations. Additionally, we explain why laser pulses of the form 
$V(t) = (n\hbar/\tau)/\cosh(t/\tau)$ do not transfer population between the levels of a two-level atom, for {\it any} choice of detuning. 

The most immediate open question is what specific multi-soliton solutions the $\hat{H}_n$'s correspond to for $n>2$ for the sine-Gordon equation. (The soliton solutions generated by $\hat{H}_n$ for the nonlinear Schr\"{o}dinger equation are given in Ref. \cite{schrader1995_2221}.) 
Additionally, our analysis only contains one free-parameter $\xi$, which is not 
sufficient to fully classify the known multi-soliton solutions. Additional freedoms 
may come from different factorization energies leading to other SUSY--QM chains (similar to the analysis in Ref.\ \cite{sukumar1986_2297} for the KdV case), 
and from the inclusion of spatial-shifts to the initial conditions at every step of the SUSY--QM chains. 
 

\section*{ACKNOWLEDGMENTS}
We are grateful to Vanja Dunjko and Steven Jackson for enlightening discussions on the subject. 
This work was supported by grants from the Office of Naval Research ({\it N00014-06-1-0455}) 
and the National Science Foundation ({\it PHY-0621703} and {\it PHY-0754942}).

%

\bibliography{Nonlinear_PDEs_and_SUSY.bib}

\begin{thebibliography}{18}
\expandafter\ifx\csname natexlab\endcsname\relax\def\natexlab#1{#1}\fi
\expandafter\ifx\csname bibnamefont\endcsname\relax
  \def\bibnamefont#1{#1}\fi
\expandafter\ifx\csname bibfnamefont\endcsname\relax
  \def\bibfnamefont#1{#1}\fi
\expandafter\ifx\csname citenamefont\endcsname\relax
  \def\citenamefont#1{#1}\fi
\expandafter\ifx\csname url\endcsname\relax
  \def\url#1{\texttt{#1}}\fi
\expandafter\ifx\csname urlprefix\endcsname\relax\def\urlprefix{URL }\fi
\providecommand{\bibinfo}[2]{#2}
\providecommand{\eprint}[2][]{\url{#2}}

\bibitem[{\citenamefont{Kay and Moses}(1956)}]{kay1956_1503}
\bibinfo{author}{\bibfnamefont{I.}~\bibnamefont{Kay}} \bibnamefont{and}
  \bibinfo{author}{\bibfnamefont{H.~E.} \bibnamefont{Moses}},
  \bibinfo{journal}{J. App. Phys.} \textbf{\bibinfo{volume}{27}},
  \bibinfo{pages}{1503} (\bibinfo{year}{1956}).

\bibitem[{\citenamefont{Shabat}(1992)}]{shabat1992_303}
\bibinfo{author}{\bibfnamefont{A.}~\bibnamefont{Shabat}},
  \bibinfo{journal}{Inverse Problems} \textbf{\bibinfo{volume}{8}},
  \bibinfo{pages}{303} (\bibinfo{year}{1992}).

\bibitem[{\citenamefont{Spiridonov}(1992)}]{spiridonov1992_398}
\bibinfo{author}{\bibfnamefont{V.}~\bibnamefont{Spiridonov}},
  \bibinfo{journal}{Phys. Rev. Lett.} \textbf{\bibinfo{volume}{69}},
  \bibinfo{pages}{398} (\bibinfo{year}{1992}).

\bibitem[{\citenamefont{Witten}(1981)}]{witten1981_513}
\bibinfo{author}{\bibfnamefont{E.}~\bibnamefont{Witten}},
  \bibinfo{journal}{Nucl.\ Phys.\ B} \textbf{\bibinfo{volume}{188}},
  \bibinfo{pages}{513} (\bibinfo{year}{1981}).

\bibitem[{\citenamefont{Sukumar}(1985)}]{sukumar1985_2917}
\bibinfo{author}{\bibfnamefont{C.~V.} \bibnamefont{Sukumar}},
  \bibinfo{journal}{J. Phys. A} \textbf{\bibinfo{volume}{18}},
  \bibinfo{pages}{2917} (\bibinfo{year}{1985}).

\bibitem[{\citenamefont{Sukumar}(1986)}]{sukumar1986_2297}
\bibinfo{author}{\bibfnamefont{C.~V.} \bibnamefont{Sukumar}},
  \bibinfo{journal}{J. Phys. A} \textbf{\bibinfo{volume}{19}},
  \bibinfo{pages}{2297} (\bibinfo{year}{1986}).

\bibitem[{\citenamefont{Barclay et~al.}(1993)\citenamefont{Barclay, Dutt,
  Gangopadhyaya, Khare, Pagnamenta, and Sukhatme}}]{barclay1993_2786}
\bibinfo{author}{\bibfnamefont{D.~T.} \bibnamefont{Barclay}},
  \bibinfo{author}{\bibfnamefont{R.}~\bibnamefont{Dutt}},
  \bibinfo{author}{\bibfnamefont{A.}~\bibnamefont{Gangopadhyaya}},
  \bibinfo{author}{\bibfnamefont{A.}~\bibnamefont{Khare}},
  \bibinfo{author}{\bibfnamefont{A.}~\bibnamefont{Pagnamenta}},
  \bibnamefont{and} \bibinfo{author}{\bibfnamefont{U.}~\bibnamefont{Sukhatme}},
  \bibinfo{journal}{Phys. Rev. A} \textbf{\bibinfo{volume}{48}},
  \bibinfo{pages}{2786} (\bibinfo{year}{1993}).

\bibitem[{\citenamefont{Cooper et~al.}(1988)\citenamefont{Cooper, Khare, Musto,
  and Wipf}}]{cooper1988_1}
\bibinfo{author}{\bibfnamefont{F.}~\bibnamefont{Cooper}},
  \bibinfo{author}{\bibfnamefont{A.}~\bibnamefont{Khare}},
  \bibinfo{author}{\bibfnamefont{R.}~\bibnamefont{Musto}}, \bibnamefont{and}
  \bibinfo{author}{\bibfnamefont{A.}~\bibnamefont{Wipf}},
  \bibinfo{journal}{Annals of Physics} \textbf{\bibinfo{volume}{187}},
  \bibinfo{pages}{1} (\bibinfo{year}{1988}).

\bibitem[{\citenamefont{Gardner et~al.}(1967)\citenamefont{Gardner, Greene,
  Kruskal, and Miura}}]{gardner1967_1095}
\bibinfo{author}{\bibfnamefont{C.~S.} \bibnamefont{Gardner}},
  \bibinfo{author}{\bibfnamefont{J.~M.} \bibnamefont{Greene}},
  \bibinfo{author}{\bibfnamefont{M.~D.} \bibnamefont{Kruskal}},
  \bibnamefont{and} \bibinfo{author}{\bibfnamefont{R.~M.} \bibnamefont{Miura}},
  \bibinfo{journal}{Phys. Rev. Lett.} \textbf{\bibinfo{volume}{19}},
  \bibinfo{pages}{1095} (\bibinfo{year}{1967}).

\bibitem[{\citenamefont{Scott et~al.}(1973)\citenamefont{Scott, Chu, and
  McLaughlin}}]{scott1973_1443}
\bibinfo{author}{\bibfnamefont{A.~C.} \bibnamefont{Scott}},
  \bibinfo{author}{\bibfnamefont{F.~Y.~F.} \bibnamefont{Chu}},
  \bibnamefont{and} \bibinfo{author}{\bibfnamefont{D.~W.}
  \bibnamefont{McLaughlin}}, \bibinfo{journal}{Proc. IEEE}
  \textbf{\bibinfo{volume}{61}}, \bibinfo{pages}{1443} (\bibinfo{year}{1973}).

\bibitem[{\citenamefont{Akulin}(2006)}]{akulin2006_book}
\bibinfo{author}{\bibfnamefont{V.~M.} \bibnamefont{Akulin}},
  \emph{\bibinfo{title}{Coherent Dynamics of Complex Quantum\\ Systems}}
  (\bibinfo{publisher}{Springer}, \bibinfo{address}{Heidelberg},
  \bibinfo{year}{2006}).

\bibitem[{\citenamefont{Delone and Krainov}(1985)}]{delone_1985}
\bibinfo{author}{\bibfnamefont{N.}~\bibnamefont{Delone}} \bibnamefont{and}
  \bibinfo{author}{\bibfnamefont{V.}~\bibnamefont{Krainov}},
  \emph{\bibinfo{title}{Atoms in Strong Light Fields}}
  (\bibinfo{publisher}{Springer}, \bibinfo{address}{Heidelberg},
  \bibinfo{year}{1985}).

\bibitem[{\citenamefont{Sall'}(1982)}]{sall1982_1092}
\bibinfo{author}{\bibfnamefont{M.~A.} \bibnamefont{Sall'}},
  \bibinfo{journal}{Theor.\ Math.\ Phys.} \textbf{\bibinfo{volume}{53}},
  \bibinfo{pages}{1092} (\bibinfo{year}{1982}).

\bibitem[{\citenamefont{V.~B.~Matveev}(1991)}]{matveev1991_book}
\bibinfo{author}{\bibfnamefont{M.~A.~S.} \bibnamefont{V.~B.~Matveev}},
  \emph{\bibinfo{title}{Darboux Transformations and \\ Solitons}}
  (\bibinfo{publisher}{Springer}, \bibinfo{address}{Heidelberg},
  \bibinfo{year}{1991}).

\bibitem[{\citenamefont{Fordy}(1994)}]{Fordy_solitons}
\bibinfo{author}{\bibfnamefont{A.}~\bibnamefont{Fordy}}, in
  \emph{\bibinfo{booktitle}{Harmonic Maps and Integrable Systems}}, edited by
  \bibinfo{editor}{\bibfnamefont{A.}~\bibnamefont{Fordy}} \bibnamefont{and}
  \bibinfo{editor}{\bibfnamefont{J.}~\bibnamefont{Wood}}
  (\bibinfo{year}{1994}), pp. \bibinfo{pages}{7--28}.

\bibitem[{\citenamefont{Drazin and Johnson}(1989)}]{solitons_book}
\bibinfo{author}{\bibfnamefont{P.~G.} \bibnamefont{Drazin}} \bibnamefont{and}
  \bibinfo{author}{\bibfnamefont{R.~S.} \bibnamefont{Johnson}},
  \emph{\bibinfo{title}{Solitons: an Introduc- \\ tion}}
  (\bibinfo{publisher}{Cambridge University Press}, \bibinfo{address}{New
  York}, \bibinfo{year}{1989}).

\bibitem[{\citenamefont{Ablowitz et~al.}(1973)\citenamefont{Ablowitz, Kaup,
  Newell, and Segur}}]{ablowitz1973_125}
\bibinfo{author}{\bibfnamefont{M.~J.} \bibnamefont{Ablowitz}},
  \bibinfo{author}{\bibfnamefont{D.~J.} \bibnamefont{Kaup}},
  \bibinfo{author}{\bibfnamefont{A.~C.} \bibnamefont{Newell}},
  \bibnamefont{and} \bibinfo{author}{\bibfnamefont{H.}~\bibnamefont{Segur}},
  \bibinfo{journal}{Phys. Rev. Lett.} \textbf{\bibinfo{volume}{31}},
  \bibinfo{pages}{125} (\bibinfo{year}{1973}).

\bibitem[{\citenamefont{Schrader}(1995)}]{schrader1995_2221}
\bibinfo{author}{\bibfnamefont{D.}~\bibnamefont{Schrader}},
  \bibinfo{journal}{IEEE J.\ Quantum Electron.} \textbf{\bibinfo{volume}{31}},
  \bibinfo{pages}{2221} (\bibinfo{year}{1995}).

\end{thebibliography}

\end{document}